\documentclass[aps,prb,twocolumn,groupedaddress,showpacs,amsmath,amssymb]{revtex4-1}

\usepackage{graphicx}
\usepackage{bm}%
\usepackage{color}

\newcommand{\be}{\begin{eqnarray}}
\newcommand{\ee}{\end{eqnarray}}

\begin{document}


\title{Effects of a magnetic field on vortex states in superfluid $^3$He-B}


\author{Kenichi Kasamatsu$^1$}
\author{Ryota Mizuno$^2$}
\author{Tetsuo Ohmi$^1$}
\author{Mikio Nakahara$^{3,4}$}
\affiliation{$^1$Department of Physics, Kindai University, Higashi-Osaka, Osaka 577-8502, Japan}
\affiliation{$^2$Department of Physics, Osaka University, 1-1 Machikaneyama-cho, Toyonaka, Osaka, 560-0043, Japan}
\affiliation{$^3$Department of Mathematics, Shanghai University, 99 Shangda Road, Shanghai 200444, China}
\affiliation{$^4$Research Institute for Science and Technology, Kindai University, Higashi-Osaka, 577-8502, Japan}


\date{\today}

\begin{abstract}
Superfluid $^3$He-B possesses three locally stable vortices known as a normal-core
vortex ($o$-vortex), an A-phase-core vortex ($v$-vortex), and a double-core vortex ($d$-vortex).
In this work, we study the effects of a magnetic field parallel or perpendicular to
the vortex axis on these 
structures by solving the two-dimensional Ginzburg-Landau equation for two different sets 
of strong coupling correction. 
The energies of the $v$- and $d$-vortices have nontrivial dependence on the 
magnetic field. As a longitudinal magnetic field increases, the $v$-vortex is energetically unstable even for high pressures 
and the $d$-vortex becomes energetically most stable for all possible range of pressure. 
For a transverse magnetic field the energy of the $v$-vortex becomes lower than that of the $d$-vortex in the 
high pressure side. In addition, the orientation of the double cores in the $d$-vortex prefers to be parallel to the magnetic field at low pressures, 
while the $d$-vortex with the double cores perpendicular to the magnetic field is allowed to continuously deform into the $v$-vortex by 
increasing the pressure.
\end{abstract}



\maketitle

\section{Introduction}{\label{intro}}
Superfluid $^3$He is a typical anisotropic Fermionic superfluid consisting of 
spin-triplet $p$-wave Cooper pairs.\cite{Vollhardt} The order parameter of 
superfluid $^3$He is characterized by $3 \times 3$ complex fields, which allows 
an existence of a rich variety of topological excitations. 
An example of topological excitations is a quantized vortex. 
Quantized vortices in superfluid $^3$He involve extremely rich physical phenomena 
and this topic has been studied for decades.\cite{SalomaaRev,Krusiusrev} 

Quantized vortices in superfluid $^3$He are classified 
by the symmetry property.\cite{Salomaa,Salomaasym} 
Several types of axisymmetric vortices exist, depending on 
which discrete symmetries are preserved in the order parameter of the vortex states. 
The actual vortex structure at given temperature and pressure is obtained 
by minimizing the appropriate free energy. 
For the $^3$He-B phase, from many theoretical studies, it has been known that three types of vortices 
exist as the local minima of the free energy. 
One possible structure is known as the $o$-vortex,\cite{Ohmi,Theodorakis} which preserves the maximal 
symmetry group and its core is filled with the normal component. 
However, the $o$-vortex is not the absolute energy minimum state in the parameter region 
where the B phase is stable.\cite{Salomaa,Passvogel} 
Another axisymmetric vortex is the $v$-vortex,\cite{Salomaa,Passvogel} which has magnetic reflection symmetry 
on a plane including the vortex line. The core of the $v$-vortex is filled with 
the A-phase and the $\beta$-phase components. In experiments, the $v$-vortex 
exists in high-temperature and high-pressure region in the $p$-$T$ phase diagram of superfluid $^3$He-B.\cite{Hakonen} 
For low-temperature and low-pressure region, however, the axisymmetry of the $v$-vortex is spontaneously 
broken into the double vortex-core structure, known as the $d$-vortex. 
Thuneberg revealed this double-core structure by direct numerical minimization of the 
Ginzburg-Landau (GL) free energy functional.\cite{Thuneberg1,Thuneberg2} 
The $d$-vortex can be interpreted as two half-quantum vortices bound together by the planar phase.\cite{Thuneberg1,Thuneberg2,Salomaa2} 
The nonaxisymmetric feature of the $d$-vortex has been experimentally observed by Kondo \textit{et al}. \cite{Kondo} 
through the measurement of a new Goldstone mode associated with the spiral twisting of the anisotropic core. 
The vortex core structure has been also calculated based on the quasiclassical Eilenberger equation.\cite{Fogelstrom,Tsutsumi,Silaev} 
This method is applicable to the weak-coupling regime, although it cannot explain the 
pressure dependence of the vortex structure in the strong-coupling regime. 

The purpose of this paper is to analyze the influence of a magnetic field 
on the stability of the aforementioned vortices. We calculate the energy of the vortices 
using the GL free energy, and construct the phase diagram of vortices in the pressure and 
magnetic-field-strength plane. 
Here, the effect of a magnetic field is included through the quadratic term 
in the GL free energy. The energetic stability is nontrivial because the magnetic field 
suppresses some order parameter components in the core of the $v$- and $d$-vortices.
To obtain the reliable results of the vortex energy, it is important to use the newly-proposed values of 
the strong coupling correction for the $\beta$-parameters in the bulk GL free energy.\cite{Saulscom} 
To this end, we adapt two different data sets. One is the Sauls-Serene values (Set I),\cite{Sauls} 
which was employed in the seminal paper by Thuneberg.\cite{Thuneberg1,Thuneberg2} 
The other is obtained by optimizing various experimental data \cite{Choi} (Set II). 
We find that, although both parameter sets give qualitatively similar vortex phase diagrams without a magnetic field, 
the Set II provides it closer to the experimental observations.  
When a magnetic field parallel to the vortex axis is turned on within the range where the bulk B-phase is thermodynamically stable, 
the $d$-vortex is the most stable structure throughout the pressure range. 
A magnetic field perpendicular to the vortex axis stabilizes the $v$-vortex; as 
the field strength increases, the stable region of the $v$-vortex extends to the lower pressure region. 
It also breaks the rotational degeneracy with respect to the symmetry axis of the $d$-vortex.

This paper is organized as follows. In Sec.~\ref{formu}, we outline the GL free energy 
required to analyze the vortex structure in superfluid $^3$He-B and list two sets of parameters in the energy functional. 
Section \ref{bulkenergysec} briefly reviews the bulk energy of superfluid $^3$He 
and Sec. \ref{nmeri} describes the prescription of the numerical calculation. 
In Sec.~\ref{without}, we show the numerical results for the case without the magnetic 
field, which is compared with those of the previous literature to confirm the validity of our calculation. 
Section \ref{mag} is the main part of our paper, where we show the numerical analysis of vortices 
in the presence of a magnetic field parallel (Sec.~\ref{Hzresults}) and perpendicular (Sec.~\ref{Hxresults}) to the vortex axis. 
Section~\ref{concle} is devoted to conclusion.

\section{Ginzburg-Landau theory of vortex states in superfluid $^3$He}
We first introduce a formulation based on the GL free energy functional for the order parameter of superfluid $^3$He. 
After describing the numerical procedure of the energy minimization, 
we analyze vortices in the B-phase without a magnetic field and 
compare the results with those of the previous work to justify our calculation. 
We repeat the calculation with the parameters Set I and Set II 
for the strong coupling corrections and reproduce the results 
in Thuneberg \cite{Thuneberg1,Thuneberg2} for Set I. The results for Set II have not been reported before. 

\subsection{Ginzburg-Landau Free Energy}\label{formu} 
The order parameter of superfluid $^3$He is given by a gap function $\hat{\Delta}$, which 
is a symmetric $2 \times 2$ matrix in spin space.\cite{Vollhardt,SalomaaRev} 
By introducing the $\bm{d}$-vector $\bm{d} = (d_x,d_y,d_z)$, the gap function 
can be written as $\hat{\Delta} = i d_{\mu} \sigma_\mu \sigma_y$, 
where $\bm{\sigma} = (\sigma_x,\sigma_y,\sigma_z)$ is the Pauli matrix and 
summation over repeated indices is understood. 
By incorporating the representation $d_\mu = A_{\mu i} \hat{p}_{i}$ with 
unit vector $\hat{\bm{p}}=(\hat{p}_{x}, \hat{p}_{y}, \hat{p}_{z})$ of the momentum on the Fermi surface, we write 
\begin{align}
\hat{\Delta}(\hat{\bm{p}})=iA_{\mu i}\sigma_{\mu}\sigma_y\hat{p}_i,\quad (\mu,i=x,y,z).
\end{align}
Thus, superfluid $^3$He is characterized by complex tensor $(A_{\mu i})$ 
inherent in the $p$-wave ($L=1$) and spin-triplet ($S=1$) pairing, 
where $\mu$ and $i$ refer to the spin and the orbital indices, respectively. 

The GL theory is valid when we confine ourselves within the situation 
near the superfluid transition temperature $T \lesssim T_c$.
By using $A_{\mu i}$, the bulk free energy density in the GL expansion takes the form
\begin{align}
f_\text{B} = & -\alpha(t) A^*_{\mu i}A_{\mu i}+\beta_1A^*_{\mu i}A^*_{\mu i}A_{\nu j}A_{\nu j} \nonumber \\
&+\beta_2 A^*_{\mu i}A_{\mu i}A^*_{\nu j}A_{\nu j}
+\beta_3 A^*_{\mu i}A^*_{\nu i}A_{\mu j}A_{\nu j} \nonumber \\
&+\beta_4 A^*_{\mu i}A_{\nu i}A^*_{\nu j}A_{\mu j}
+\beta_5 A^*_{\mu i}A_{\nu i}A_{\nu j}A^*_{\mu j}, \label{GLbulkene}
\end{align}
which is invariant under separate spin and real space rotations 
in addition to the gauge transformation, namely U(1)$\times$SO$^{(S)}$(3)$\times$SO$^{(L)}$(3).
The coefficient $\alpha(t)$ of the second order term has a temperature dependence 
$\alpha(t) = \alpha_0 t$ with a constant $\alpha_0$ and a small parameter $t = 1 - T / T_c $. 

\begin{figure}[ht]
\centering
\includegraphics[width=0.75\linewidth]{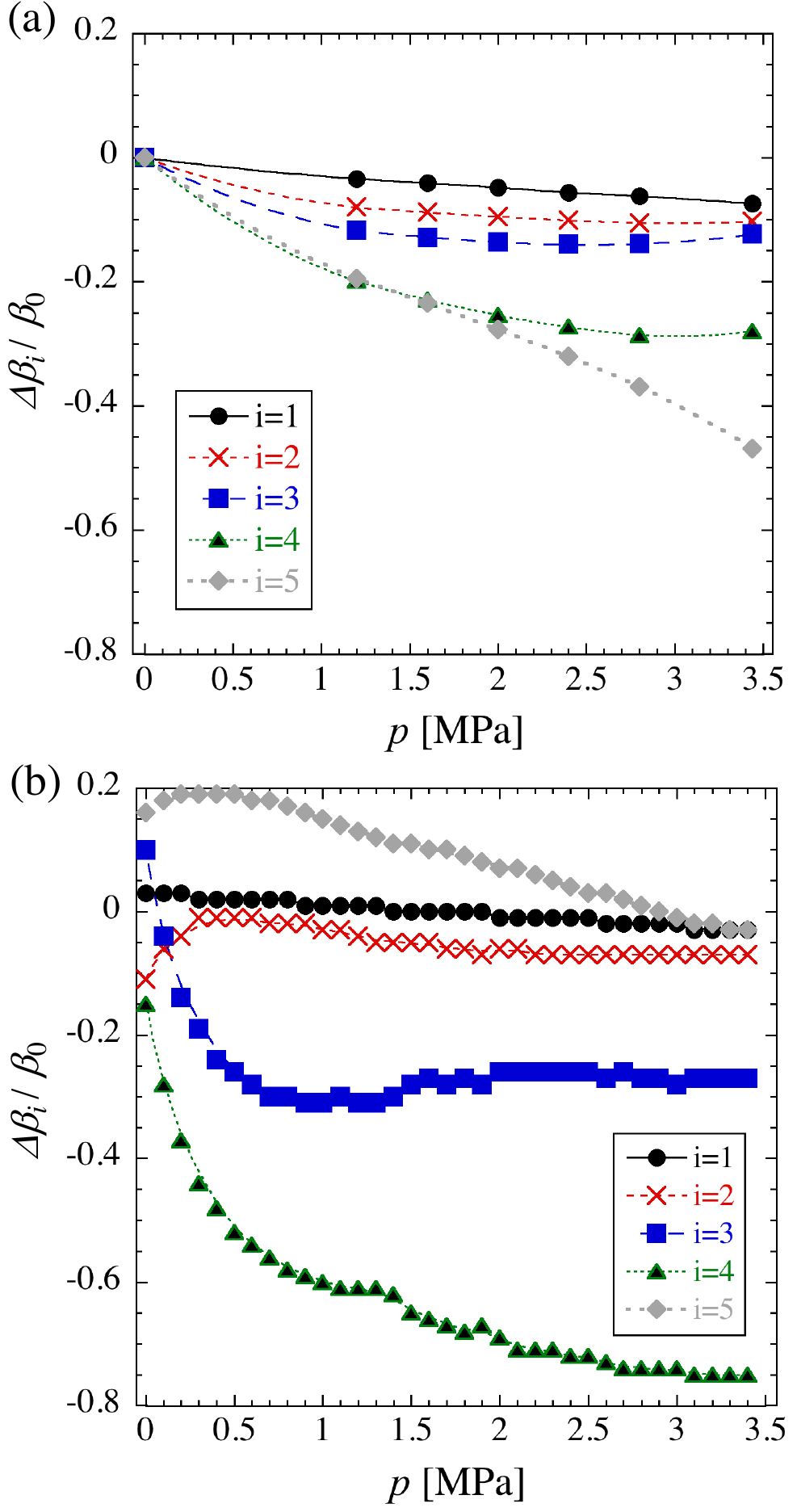} 
\caption{(Color online) Strong coupling correction $\Delta \beta_i$ ($i$=1-5). The data points are taken from Ref.~\cite{Sauls} [Set I] (a) and Ref.~\cite{Choi} [Set II] (b), and 
the curves represent the interpolating functions.}
\label{betav}
\end{figure}
We consider the pressure range $0 \leq p \leq 3.44$ MPa, 
beyond which the system solidifies. 
The fourth order contribution to the bulk free energy has five terms with coefficients $\beta_j$. 
In the weak-coupling limit at zero pressure, these coefficients satisfy 
the relation
\begin{equation}
-2\beta_1^{\mathrm{wc}}=\beta_2^{\mathrm{wc}}=\beta_3^{\mathrm{wc}}
=\beta_4^{\mathrm{wc}}=-\beta_5^{\mathrm{wc}} \equiv 2 \beta_0, 
\label{weakcouplingbeta}
\end{equation}
where $\beta_0 = 7 \zeta(3) N (0)/[120 \pi^2 (k_B T_c)^2] $ with the density of states at the Fermi surface 
per spin $N(0)$ and the Riemann zeta function $\zeta(3)$.
The pressure dependence can be included by the strong-coupling corrections for $\beta_j$, 
which are written as $\Delta \tilde{\beta}_j = \Delta \beta_j /\beta_0= (\beta_j-\beta_j^{\mathrm{wc}})/\beta_0$.
However, the exact values of the correction as functions of the pressure have not been known. 
Here, we use two data sets. 
One is the theoretical evaluation by Sauls and Serene.\cite{Sauls} 
In their paper, the values of $\Delta \tilde{\beta}_j$ for six values of pressure are given. 
In our work, these points and $\Delta \tilde{\beta}_j = 0$ at $p=0$, where $\beta_j(p=0)$ 
coincide with the weak-coupling values of Eq.~\eqref{weakcouplingbeta}, 
are interpolated through the 4-th order polynomials 
$\Delta \tilde{\beta}_j = \sum_{k=1}^{4} A_{j}^{(k)} p^k$ as shown in Fig.~\ref{betav} (a), where $p$ [MPa] is pressure . 
The list of the fitting parameters $A_{j}^{(k)}$ are shown in Table~\ref{betacoefficient1}. 
We refer to this set of $\beta_j$ as ``Set I".
\begin{table}
\caption{\label{betacoefficient1} List of the coefficients for the fitting of the correction of $\beta$ according 
to the paper by Sauls and Serene.\cite{Sauls} Write ``Set I".}
\begin{ruledtabular}
\begin{tabular}{ccccc}
$j$ & $A_j^{(1)}$ & $A_j^{(2)}$ & $A_j^{(3)}$ & $A_j^{(4)}$  \\
\hline
1 & $-0.041318$ & $0.015586$ & $-0.0043854$ & $0.00044482$ \\
2 & $-0.13306$ & $0.081867$ & $-0.025704$ & $0.0030876$  \\
3 & $-0.18713$ & $0.10679$ & $-0.030986$ & $0.0037026$  \\
4 & $-0.28746$ & $0.14638$ & $-0.043984$ & $0.0054785$ \\
5 & $-0.24049$ & $0.090797$ & $-0.023003$ & $0.0015738$  \\
\end{tabular}
\end{ruledtabular}
\end{table}

\begin{table*}
\caption{\label{betacoefficient2} List of the coefficients for the fitting of the correction of $\beta_j$ according to 
the paper by Choi et al.\cite{Choi} Write ``Set II".}
\begin{ruledtabular}
\begin{tabular}{ccccccccccc}
$j$ &  $B_j^{(0)}$ & $B_j^{(1)}$ & $B_j^{(2)}$ & $B_j^{(3)}$ & $B_j^{(4)}$ & $B_j^{(5)}$ & $B_j^{(6)}$ & $B_j^{(7)}$ & $B_j^{(8)}$ & $B_j^{(9)}$ \\
\hline
1 & $0.030472$ & $0.018587$ &$ -0.34839$ & $1.0919$ &  $-1.66$ & $1.4058$ & $-0.69863$ & $0.2027$ & $-0.031801$  & $0.0020847$\\
2 & $-0.11014$ & $0.61315$ & $-1.4457$ & $1.8258$  &  $-1.5064$ & $0.84442$ & $-0.31656$ & $0.075439$ & $-0.010271$ &$0.00060559$ \\
3 & $0.10094$ & $-1.7982$ & $4.1219$ & $-6.3112$  &   $6.4351$  & $-4.2344$ & $1.7664$ & $-0.45099$ & $0.06427$ & $-0.0039131$\\
4 & $-0.15024$ & $-1.5428$ & $2.8253$ & $-3.82$ &  $3.9311$ & $-2.858$ & $1.3449$ & $-0.3822$ & $0.059274$  & $-0.0038433$\\
5 & $0.16064$ & $0.25229$ & $-0.79652$ & $1.4591$ &  $-1.9474$ & $1.655$ & $-0.84977$ & $0.25472$ &  $-0.04103$ & $0.0027444$ \\
\end{tabular}
\end{ruledtabular}
\end{table*}
The other is described in Choi \textit{et al}.,\cite{Choi} which reports a new data 
of the $\beta_i$-values as functions of the pressure based on past 
experiments. The thermodynamic properties of 
superfluid $^3$He have been better accounted for by the analysis with 
this new $\beta_j$-values.\cite{Wiman} 
We also use the values of $\beta_j$ following Ref.~\cite{Choi}; 
numerical fitting of the data with the 9-th order polynomials 
$\Delta \tilde{\beta}_j = \sum_{k=0}^{9} B_{j}^{(k)} p^k$ is shown in Fig.~\ref{betav} (b), 
where the coefficients $B_{j}^{(k)}$ are listed in Table~\ref{betacoefficient2}. 
We refer to this set of $\beta_j$ as ``Set II''.
The obvious differences between Figs.~\ref{betav}(a) and \ref{betav}(b) are that the values of $\beta_i$ at $p=0$ are 
slightly shifted from the weak-coupling values, and the magnitude of the correction is relatively large 
for $\beta_3$ and $\beta_4$ in Fig.~\ref{betav}(b); especially 
$\Delta \beta_5$ gives a positive correction in contrast with the Set I case. 
This parameter set energetically favors the emergence of the A-phase component 
in the bulk phase diagram compared to the case of Set I, as seen in Ref.\cite{Wiman} and discussed below. 

The gradient energy in the GL expansion is given by
\begin{align}
f_\text{G} = K _1 \partial _i A _{\mu j} \partial _i A _{\mu j} ^* +
K _2 \partial _i A _{\mu i} \partial _j A _{\mu j} ^* 
+K _3 \partial _i A _{\mu j} \partial _j A _{\mu i} ^*.
\label{gradene}
\end{align}
The coefficients $K_i$ satisfy $K _1 = K _2 = K _3 = K$ in the weak-coupling limit, which we 
employ in the rest of this paper for simplicity. By comparing the gradient term and the 
first term of the bulk energy [Eq.~(\ref{GLbulkene})], the coherence length is defined as
\begin{align}
\xi(t) = \sqrt{\frac{K}{ \alpha (t)}} = \sqrt{\frac{K}{\alpha_0}} \frac{1}{\sqrt{t}}.
\end{align}
The minimum length scale of our problem is the coherence length 
at zero temperature $\xi(0) =  \sqrt{K/ \alpha_0} \sim 0.01$ $\mu$m. 

In this work, we consider the effect of a magnetic field on the vortex state. 
We take account of this effect through the second order magnetic free energy 
given by 
\begin{align}
f_\text{M}&=g_m H_{\mu}A_{\mu i}^*H_{\nu}A_{\nu i},
\label{ffield}
\end{align}
where the coefficient $g_m$ is given in the weak-coupling limit as
\begin{equation}
g_m^\mathrm{wc} = \frac{7 \zeta(3) N(0) (\gamma \hbar)^2}{48 [(1+F_0^a) \pi k_B T_c]^2} 
\end{equation} 
with the gyromagnetic ratio $\gamma = -2.04 \times 10^5$ /(mT$\cdot$sec) and the Landau parameter $F_0^a = -0.695$ at zero pressure. 
The pressure dependence of $g_m$ was also reported in Ref.~\cite{Choi}, where 
$g_m$ is close to its weak-coupling value $g_m^\mathrm{wc}$ in $0 \leq p \leq 3.44$ MPa. 
Thus, we assume $g_m = g_m^\mathrm{wc}$ in the following calculation. 
The characteristic strength of the magnetic field is $H_{0} = \sqrt{\alpha/g_m^\mathrm{wc}} \simeq 0.79 \sqrt{t} $ [T]. 
We ignore the first order contribution of the magnetic field to the GL free energy, since 
it is sizable at a strong magnetic field $\sim 1$ T which 
stabilizes the A$_1$ phase, while our interest is under a weaker magnetic field on the order of 100 mT. 
Also, we drop the contribution of the dipole energy which provides negligible contribution to our free energy. 

\subsection{Bulk energy density} \label{bulkenergysec}
To introduce the basic scales in our problem, let us look at the 
equilibrium property of the bulk $^3$He-B phase without a magnetic field. 
The B-phase order parameter is written as $A^\text{(B)} = \Delta_\text{B} R(\hat{\bm{n}}, \phi)$, 
where $\Delta_\text{B}$ is the bulk gap amplitude and $R(\hat{\bm{n}}, \phi)$ is the rotation matrix 
with an angle $\phi$ around the axis $\hat{\bm{n}}$; the bulk energy is degenerate 
with respect to $\hat{\bm{n}}$ and $\phi$. Choosing $\phi = 0$, we have 
$A^\text{(B)} = \Delta_\text{B} \hat{I}$ with the $3 \times 3$ unit matrix $\hat{I}$.
The energy density for the bulk $^3$He-B phase is given by 
\begin{equation}
f_\text{B}^0 = - 3 \alpha \Delta_\text{B}^2 + ( 9 \beta_{12} + 3 \beta_{345} ) \Delta_\text{B}^4, 
\end{equation}
where $\beta_{ij} = \beta_i + \beta_j$ and 
$\beta_{ijk} = \beta_i + \beta_j + \beta_k$.
Minimization of this energy density leads to the bulk gap 
\begin{equation}
\Delta_\text{B} = \sqrt{\frac{\alpha}{\beta'}},  \label{bulkbphaseamp}
\end{equation}
and the corresponding energy density  
\begin{equation}
f_\text{B}^0 = - \frac{3 \alpha^2}{2 \beta'} \label{bulkbphaseeneden}
\end{equation}
with $\beta' = 6\beta_{12} + 2 \beta_{345}$. 

\begin{figure}[ht]
\centering
\includegraphics[width=1.0\linewidth]{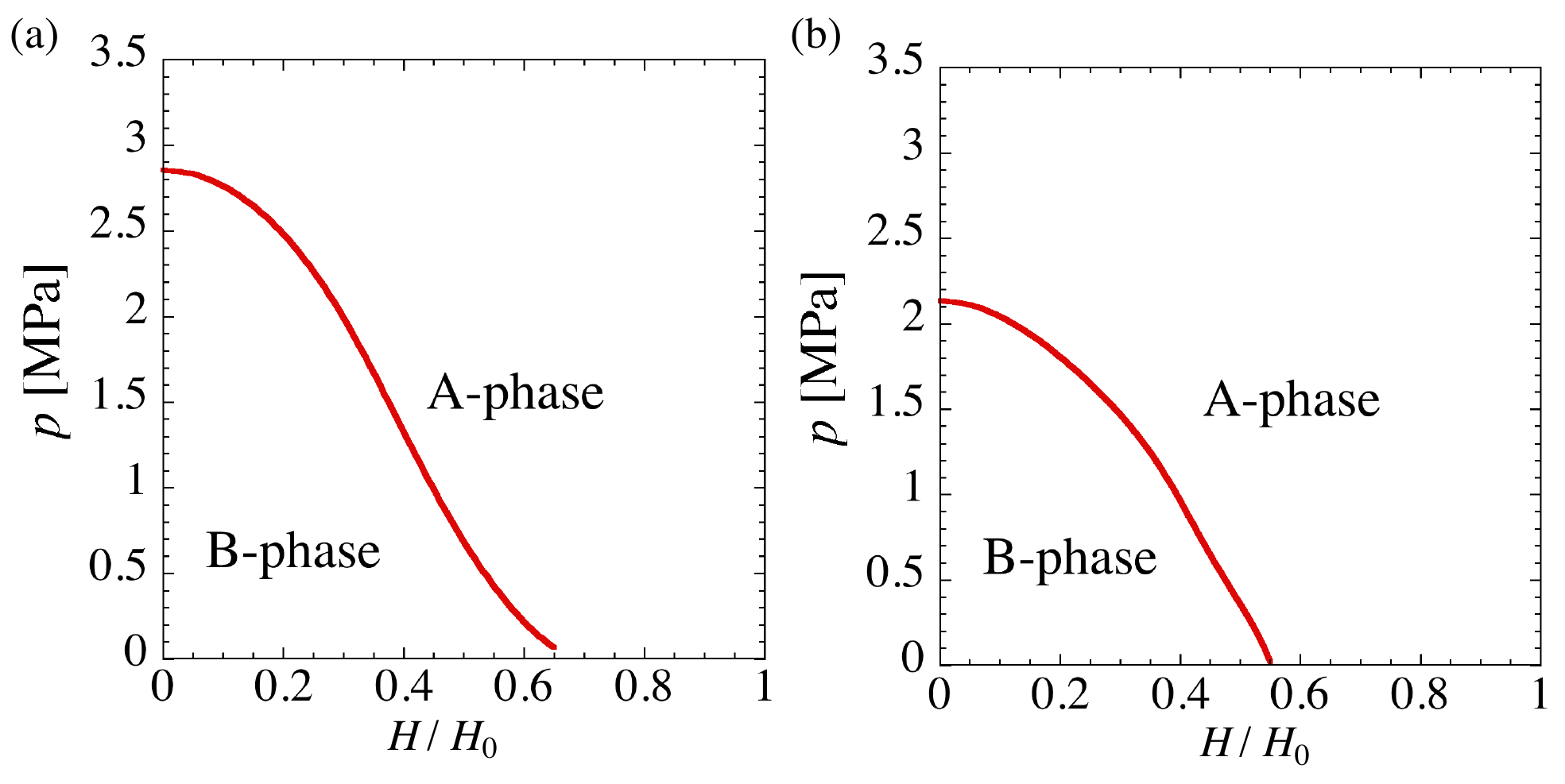} 
\caption{(Color online) Bulk phase diagram of superfluid $^{3}$He in the $p$-$H$ plane, calculated by 
the GL free energy, with Set I (a) and Set II (b) for strong-coupling corrections. }
\label{phasediabulk}
\end{figure}

In this work, a magnetic field $\bm{H} = H \bm{e}_i$ is applied along the direction 
which is parallel $(i=z)$ and perpendicular $(i=x)$ to the vortex axis. We refer to the former and latter fields 
as a longitudinal and a transverse magnetic fields, respectively.
The magnitude of $H$ is assumed to be small enough so that the system does not escape 
from the B-phase. 
To find the suitable range of $H$, we calculate the bulk free energy in the magnetic field 
$\bm{H} = H \bm{e}_z$. The order parameter of the B-phase is modified as 
\begin{align}
A^\text{(B)} =  \left( 
\begin{array}{ccc}
\Delta_{\perp} & 0 & 0 \\
0 & \Delta_{\perp} & 0 \\ 
0 & 0 & \Delta_{\parallel} 
\end{array}\right), 
\end{align}
where 
\begin{align}
\Delta_{\perp} &= \Delta_\text{B} \sqrt{ 1+ \frac{\beta_{12} g_m}{\alpha \beta_{345}} H^2 } \nonumber \\
\Delta_{\parallel} &= \Delta_\text{B} \sqrt{ 1- \frac{(2\beta_{12} + \beta_{345}) g_m}{\alpha \beta_{345}}  H^2 },   \label{magampliB}
\end{align}
and the corresponding energy density is 
\begin{align}
f_\text{B}^0 &= - \frac{3 \alpha^2}{2 \beta'} \left[ 1 - \frac{2 g_m H^2}{3 \alpha} + \frac{(2 \beta_{12} + \beta_{345}) g_m^2 H^4}{3 \alpha^2 \beta_{345}} \right].  \label{bulkenejiba} 
\end{align}
This energy is compared with that of the bulk A-phase characterized by $A^\text{(A)} = \Delta_\text{A} \hat{\bm{d}} (\hat{\bm{m}} + i \hat{\bm{n}})$ with 
arbitrary unit vectors $\hat{\bm{d}}$, $\hat{\bm{m}}$, and $\hat{\bm{n}}$ such that $\hat{\bm{m}} \cdot \hat{\bm{n}} = 0$. 
By choosing $\hat{\bm{d}} = \hat{\bm{e}}_x$, $\hat{\bm{m}} = \hat{\bm{e}}_x$, and $\hat{\bm{n}} = \hat{\bm{e}}_y$, the 
energy does not depend on the magnetic field; the order parameter and energy density are then given by 
\begin{align}
\Delta_\text{A} = \sqrt{ \frac{\alpha}{4 \beta_{245}} } , \quad
f_\text{A}^0 = - \frac{\alpha^2}{4 \beta_{245}}.
\end{align}
By comparing the energies $f_\text{B}^0$ and $f_\text{A}^0$, we obtain the bulk phase diagram in the $p$-$H$ plane as 
shown in Fig.~\ref{phasediabulk}. We see qualitatively similar phase diagrams for both data sets of the strong-coupling correction, 
where the B-phase is stable for a low pressure and a low magnetic field. The area of the stable region of the B-phase for Set II 
is smaller than that for Set I. Below we confine ourselves within the range  $H \lesssim 0.4 H_{0}$ 
in which the bulk B-phase is still local energy minimum for high pressure. 

\subsection{Numerics and boundary condition}\label{nmeri}
We compute the order parameters of vortex states by minimizing the total free energy,
\begin{align}
F =  \int d \bm{r} (f_\text{B} + f_\text{G} + f_\text{M})
\end{align}
with the vortex boundary condition. To this end, we evolve the order parameter through the relation 
\begin{equation}
\frac{\partial A_{\mu i}}{\partial \tau} = - \frac{\delta F}{\delta A_{\mu i}^{\ast}}  \label{tDGLe}
\end{equation}
in two-dimensional Cartesian coordinate, assuming homogeneity along the $z$-direction. 
Equation~\eqref{tDGLe} is known as the time-dependent GL equation; from a suitable initial state, 
the order parameter converges to the solution of the energy minimum by time-evolving 
Eq.~\eqref{tDGLe}. Here, $\tau$ is called the imaginary time. 
Axisymmetry cannot be assumed \textit{a priori} because the $d$-vortex breaks it spontaneously. 

We use the temperature-dependent coherence 
length $\xi(t) = \sqrt{K/\alpha(t)}$, $\Delta_\text{B}$ in Eq.~\eqref{bulkbphaseamp}, and 
$|f_\text{B}^0|$ in Eq.~\eqref{bulkbphaseeneden} to scale the length, the order parameter, and the energy density, respectively. 
Then, the dimensionless order parameter and the free energy is written as $ \tilde{A}_{\mu i} = A_{\mu i} / \sqrt{\alpha(t)/\beta'}$ and
\begin{equation}
\tilde{F} \equiv \frac{F}{\left| f_\text{B}^0 \right| \xi^2(t)} = \int d \tilde{\bm{r}} \frac{2}{3} (\tilde{f}_\text{B} + \tilde{f}_\text{G} + \tilde{f}_\text{M}), 
\label{dimlessfree}
\end{equation}
respectively, where tildes denote the dimensionless variables. 
Also, by scaling the magnetic field as $H = H_{0} \tilde{H}$, the energy density in 
the right hand side of Eq.~(\ref{dimlessfree}) can be written by just replacing the coefficients 
in Eqs.~(\ref{GLbulkene}), (\ref{gradene}), and (\ref{ffield}) as 
\begin{align}
& \alpha \to 1, \quad \beta_j \to \frac{\beta_j}{\beta'} = \frac{\beta_j^\text{wc} + \Delta \beta_j}{\beta'}, \nonumber \\ 
& K \to 1, \quad g_{m} \to  \frac{g_{m}(p)}{g_{m}(0)} \approx 1.
\end{align}
Then, all physical quantities become dimensionless. 

We take a system size for numerical simulations as $\tilde{x},\tilde{y} \in [-50,50]$ with a 
$500 \times 500$ numerical grid. Although we solve Eq.~\eqref{tDGLe} 
in the $xy$-coordinate, we impose the Direchlet boundary condition for the vortex 
structure on the cylinder with the radius $\tilde{R}=50$. 
The boundary condition is taken as $\lim_{\tilde{r} \to \tilde{R}} \tilde{A}_{\mu i}(\tilde{r},\theta) = \hat{I} e^{i\theta}$, 
where $(\tilde{r},\theta)$ is the dimensionless polar coordinate. 
This means that far from the vortex core at the center, the order parameter approaches to the 
bulk value with unit phase winding. 
However, as shown by Hasegawa,\cite{Hasegawa} the asymptotic form of the 
off-diagonal components of $\tilde{A}_{\mu i}$, some of which occupy the vortex core, 
decays slowly as $\sim r^{-1}$, and we have to take 
large system size to ensure the validity of this boundary condition. 
We confirm that $\tilde{R} = 50$ is large enough to ignore the finite size effect on the vortex energy. 
Furthermore we avoid the unfavorable boundary contribution to the vortex energy by 
integrating Eq.~\eqref{dimlessfree} within the region $\tilde{R} \leq 40$ when calculating the total energy. 

\subsection{Vortex state without a magnetic field}\label{without}
\begin{figure}[ht]
\centering
\includegraphics[width=0.96\linewidth]{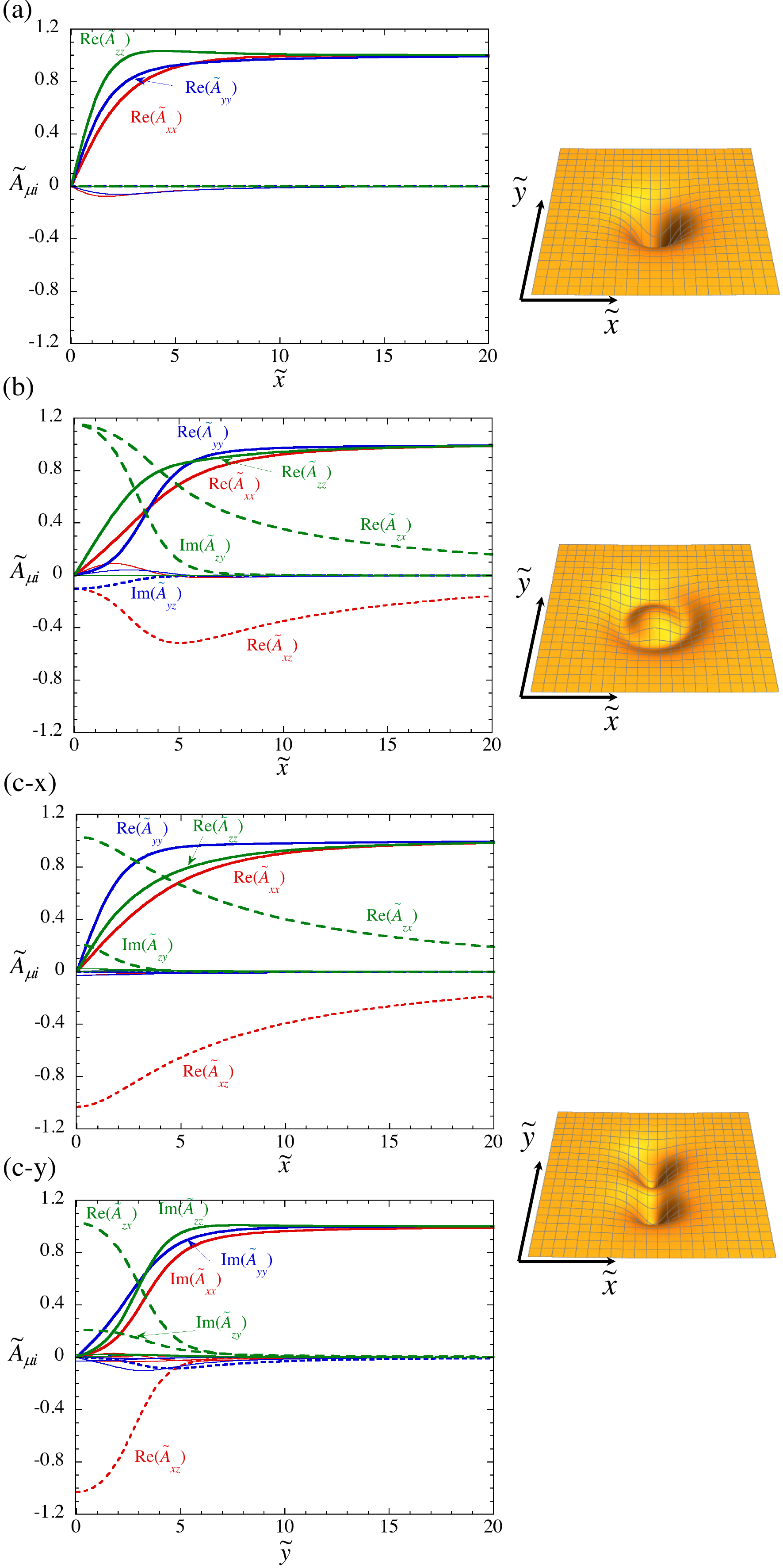} 
\caption{(Color online) Left: Typical radial profiles of the order parameters for (a) the $o$-vortex, (b) the $v$-vortex, and (c) the $d$-vortex. 
Here the strong-coupling correction Set II is employed and the pressure is set to 3.0 MPa. Because the $d$-vortex is not axisymmetric, 
we show the profiles along the $x$- and $y$-axis as (c-x) and (c-y), respectively. Diagonal components of $(\tilde{A}_{\mu i})$ are shown 
in solid curves, while off-diagonal ones in dashed curves. Right: Corresponding 2D profiles of the pair density $\sum_{\mu,i} |\tilde{A}_{\mu i}|^2$.}
\label{vorprofilezero}
\end{figure}
\begin{figure}[ht]
\centering
\includegraphics[width=0.75\linewidth]{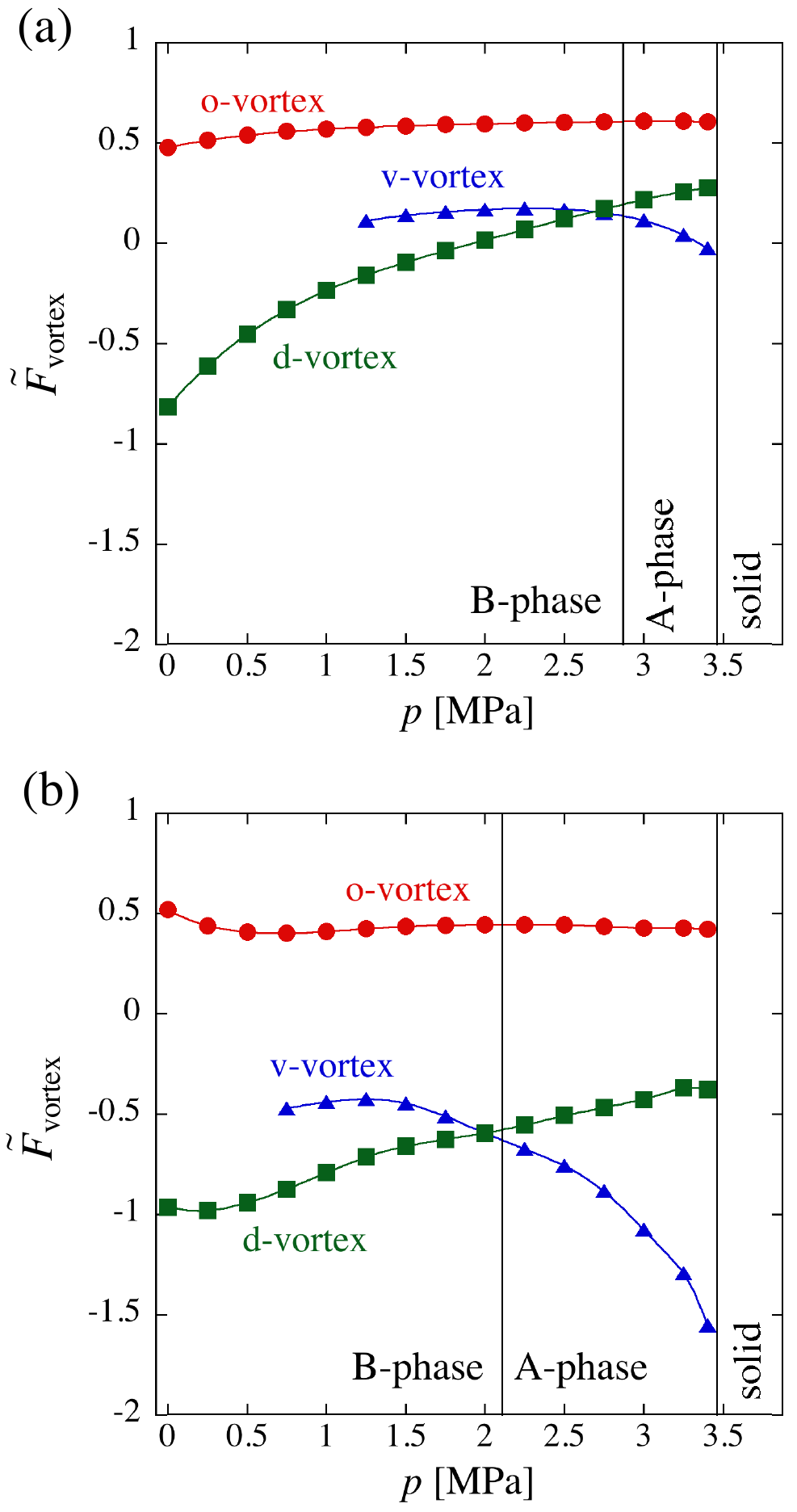} 
\caption{(Color online) Free energies of the three vortex states without a magnetic field as functions 
of the pressure $p$ for Set I (a) and Set II (b). The energies of the $o$-, $v$-, and $d$- 
vortices are shown by curves with red circles, blue triangles, and green squares, respectively. The bulk A-B phase 
boundary is taken from Fig.~\ref{phasediabulk}.}
\label{vorenezero}
\end{figure}

Here, we briefly review the vortex state without a magnetic field. Typically, there are three types of solutions, 
known as the $o$-vortex, the $v$-vortex, and the $d$-vortex. Thuneberg analyzed these vortex states and calculated their 
energies within the GL theory with $\beta$-values obtained from Set I.\cite{Thuneberg1,Thuneberg2} 
Here, we calculate them not only with Set I but also with Set II, obtaining qualitatively similar results for 
both data sets; the typical spatial profiles of the order parameters for Set II are shown in Fig.~\ref{vorprofilezero} 
and the vortex energy as a function of $p$ is shown in Fig.~\ref{vorenezero}. 
The $o$-vortex has an axisymmetric structure and the core is not filled with 
any superfluid components.  The $v$-vortex is also axisymmetric but has a core filled mainly with the A-phase components $\text{Re}(\tilde{A}_{zx})$ 
and $\text{Im}(\tilde{A}_{zy})$, whose amplitudes take the same value at the origin. There are also minor components 
$\text{Re}(\tilde{A}_{xz})$ and $\text{Im}(\tilde{A}_{yz})$ filling the vortex core, which are known as the $\beta$-phase. 
The fraction of the A-phase component increases as the pressure increases. 
The $d$-vortex is a non-axisymmetric vortex as seen in Figs.~\ref{vorprofilezero}(c-x) and (c-y), 
where the profiles are different along the $\tilde{x}$-direction and $\tilde{y}$-direction. 
The core components are $\text{Re}(\tilde{A}_{zx})$ and $\text{Re}(\tilde{A}_{xz})$, corresponding to the planar phase characterized by 
$\tilde{A}= \tilde{\Delta}_\text{p} R(\hat{\bm{n}},\phi) (I - \hat{\bm{w}} \hat{\bm{w}})$ with $\hat{\bm{n}} = \hat{\bm{w}} = \bm{e}_y$ and $\phi=\pi/2$.\cite{Thuneberg2} 
In this case, there appear double vortex cores along the $y$-direction. 
The vortex energy is calculated by subtracting $ \tilde{F}_\mathrm{bulk}$ and $\tilde{F}_\mathrm{hyd}$ 
from the total free energy $\tilde{F}$ in Eq.~\eqref{dimlessfree} as 
\begin{equation}
\tilde{F}_\text{vortex} = \tilde{F} - \tilde{F}_\mathrm{bulk} - \tilde{F}_\mathrm{hyd}.  
\end{equation}
Here, $\tilde{F}_\mathrm{bulk} = 1 \times S$ is the bulk free energy
with the system area $S$ and $F_\mathrm{hyd} = \int d \mathbf{r} \sum_{k} \rho_k v_k^2/2$ is a 
hydrodynamic kinetic energy caused by vortex flow. The energy $F_\mathrm{hyd}$ 
can be calculated by combining the relation $j_k = \rho_{k} v_{sk}$ ($k=x,y,z$) for the superfluid 
density $\rho_k$ and the current density 
\begin{align}
j_k = \frac{4 m_3 K}{\hbar} \textrm{Im} \left[  A_{\mu k}^{\ast} \partial_j A_{\mu j} + A_{\mu j}^{\ast} \partial_k A_{\mu j} 
+ A_{\mu j}^{\ast} \partial_j A_{\mu k} \right].
\end{align}
The dimensionless form of $F_\mathrm{hyd}$ is 
$\tilde{F}_\mathrm{hyd} = (20\pi/3)\ln(R/\xi(t))$ with the radius $R$ of the cylinder, where the lower bound of the integral has been taken as $\xi(t)$. 

For Set I, we reproduce the results of Thuneberg \cite{Thuneberg1,Thuneberg2} as seen in 
Fig.~\ref{vorenezero}(a) and confirm the validity of our calculation. The $o$-vortex is always the highest energy configuration.
For low pressures, the $d$-vortex is the most stable vortex state, while 
the $v$-vortex is most stable above $p \simeq 2.7$ MPa. 
In our two-dimensional calculation without axisymmetry, the $v$-vortex is energetically unstable for low pressures 
below 1.25 MPa and it decays to the $d$-vortex. The similar behavior has also been mentioned by 
Thuneberg.\cite{Thuneberg2} 

Figure~\ref{vorenezero}(b) shows the energy of the vortex states for Set II. 
Although the result qualitatively agrees with that in Fig.~\ref{vorenezero}(a), we confirm quantitative differences. 
Especially, the critical pressure for the transition between the $v$-vortex and the $d$-vortex is remarkably 
reduced to $\approx 2.0$ MPa, compared to that of $\approx 2.7$ MPa seen in Fig.~\ref{vorenezero}(a). 
Thus, the results for Set II explains the experimental observation of the vortex-core transition 
\cite{Krusiusrev,Thuneberg1} for $T \simeq T_c$ more accurately than those for Set I.
Note that the transition pressure of the $v$- and the $d$-vortices is close to the boundary of the bulk
AB transition for both Set I and Set II.

\section{Vortex state under a magnetic field}\label{mag}
In this section, we consider the effect of a magnetic field $\bm{H}$ on the vortex states. 
We address two situations: one with the longitudinal magnetic field $\bm{H} = H \bm{e}_z$, and the other 
with the transverse magnetic field $\bm{H} = H \bm{e}_x$. 

\subsection{Longitudinal magnetic field}\label{Hzresults}
Here we consider the vortices under a magnetic field along the $z$-axis. 
The effect of the axial magnetic field can be understood by examining the free energy 
functional. The quadratic magnetic-field term Eq.~\eqref{ffield} for $\bm{H} = H \bm{e}_z$ 
is written as $\tilde{f}_\text{M} = \tilde{H}^2 \sum_i |\tilde{A}_{zi}|^2$. Thus, the population of 
the core components whose spin is along the $z$-direction in the $v$- and $d$-vortices 
should be suppressed to reduce the energy cost. Then, it is not trivial to tell which vortex 
state has the lowest energy. 

To obtain the vortex solution under the magnetic field, 
we should fix the boundary condition, i.e., the proper choice of $R(\hat{\bm{n}},\phi)$ at the boundary. 
In the bulk region, by taking account of the quadratic magnetic energy and the dipole-dipole energy, the 
$\hat{\bm{n}}$ should be parallel to $\bm{H}$, since the combination of these energies gives an energy contribution
$\propto - (\hat{\bm{n}} \cdot \bm{H})^2$.\cite{Thuneberg2} Also, the dipole-dipole energy fixes $\phi = \arccos(-1/4)$. 
Since the vortices are along the $z$-axis in our problem, the vortex structures including the core components as well as the 
vortex energies are not affected by the rotation $R(\hat{\bm{z}},\phi)$ of the order parameter at the boundary. 
Thus, we set the bulk amplitude with $\phi = 0$ as the Direchlet boundary condition at $r=R$, and write the order parameter at the boundary as 
\begin{align}
\tilde{A}^\text{(B)} =  \left( 
\begin{array}{ccc}
\tilde{\Delta}_{\perp} & 0 & 0 \\
0 & \tilde{\Delta}_{\perp} & 0 \\ 
0 & 0 & \tilde{\Delta}_{\parallel} 
\end{array}\right) e^{i \theta},
\end{align}
where $\tilde{\Delta}_{\perp}$ and $\tilde{\Delta}_{\parallel}$ are given by Eqs.~\eqref{magampliB}. 
The vortex energy $\tilde{F} - \tilde{F}_\mathrm{bulk} - \tilde{F}_\mathrm{hyd}$ is calculated in a similar way as that described in Sec.~\ref{without}. 
Here, $\tilde{F}_\text{bulk}$ is evaluated by the spatial integral of Eq.~\eqref{bulkenejiba} and 
$\tilde{F}_\text{hyd} = (4\pi/3) (4 \tilde{\Delta}_{\perp}^2 +  \tilde{\Delta}_{\parallel}^2) \ln (R/\xi(t))$. 

\begin{figure}[ht]
\centering
\includegraphics[width=1.0\linewidth]{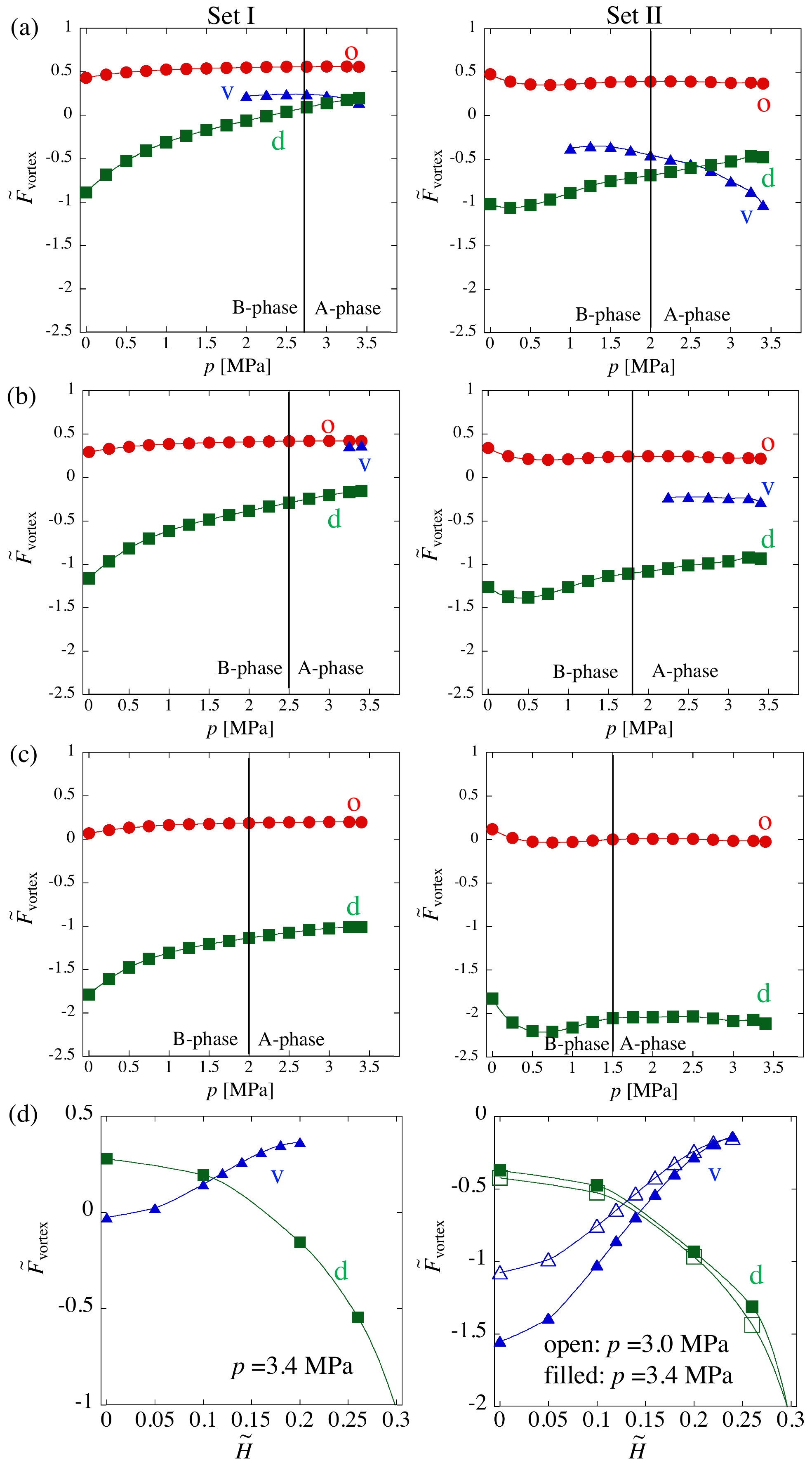} 
\caption{(Color online) Free energies of the three vortex states as functions 
of pressure $p$ for Set I (left panels) and Set II (right panels). The magnitude 
of the magnetic field is (a) $\tilde{H} = 0.1$, (b) $\tilde{H} = 0.2$, (c) $\tilde{H} = 0.3$. 
The energies of the $o$-, $v$-, and $d$-vortices are shown by curves with red circles, blue triangles, and green squares, respectively. The bulk A-B phase 
boundary is taken from Fig.~\ref{phasediabulk}. The figures (d) show the free energies of $v$- and $d$-vortices 
as functions of $\tilde{H}$ at high pressure region; we show the plots with $p=3.0$ MPa for Set I, and with $p=3.0$ and 3.4 MPa for Set II. }
\label{vorenergymag}
\end{figure}

Figure~\ref{vorenergymag} shows the vortex free energy $\tilde{F}_\text{vortex}$ 
of the three vortex states 
as a function of the pressure for several $\tilde{H}$. The main feature of Fig.~\ref{vorenergymag} for both sets of the 
strong-coupling correction is that the $d$-vortex is the most preferable structure under the magnetic field. 
The $v$-vortex has a local minimum only in a high-pressure region, which shrinks and 
eventually disappears as the magnetic field strength increases; the initial $v$-vortex relaxes 
to the $d$-vortex through the evolution of Eq.~\eqref{tDGLe} when $v$-vortex is unstable. 
The $o$-vortex is always the highest energy state, being metastable in all situations. 
Although we solved Eq.~\eqref{tDGLe} with other initial configurations, 
the final states of the imaginary time evolution are always one of the three vortices. 

\begin{figure}[ht]
\centering
\includegraphics[width=1.0\linewidth]{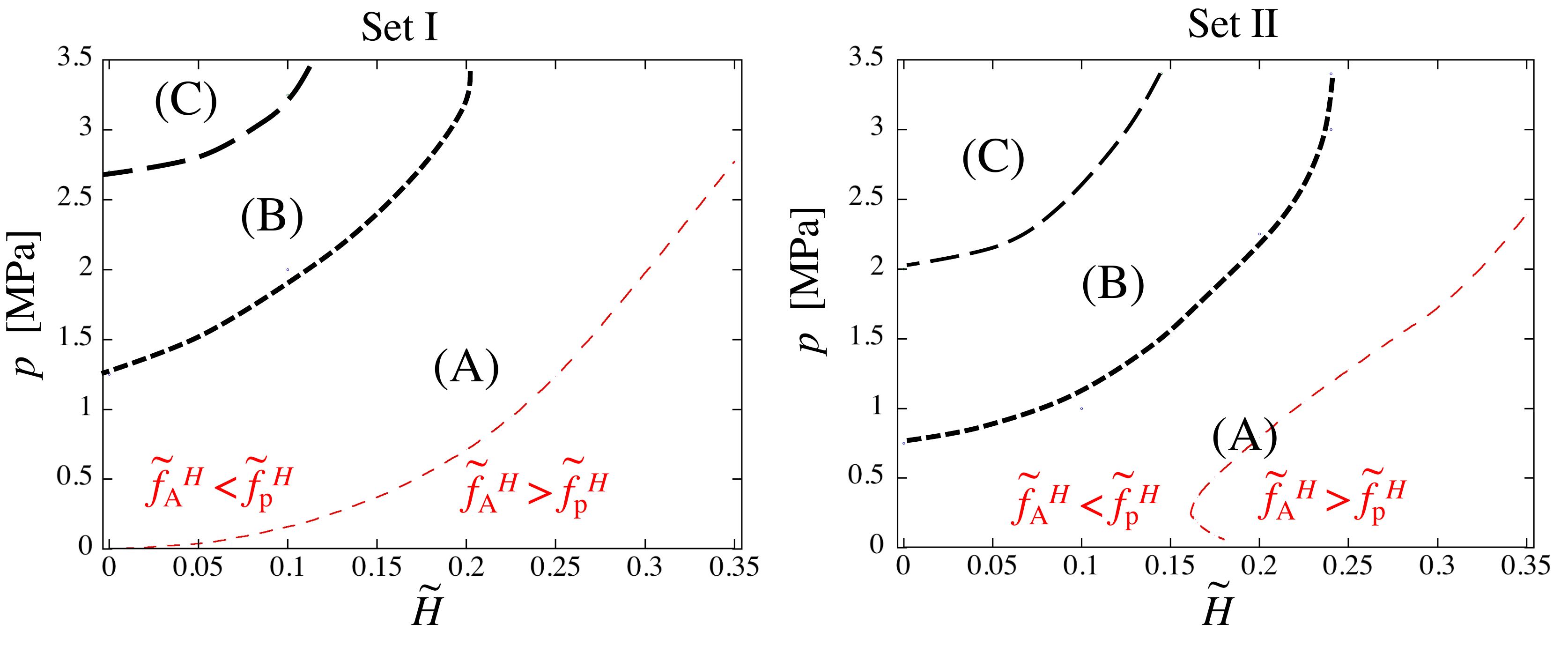} 
\caption{(Color online) $p$-$H$ phase diagram of the stable $v$- and $d$-vortex states 
for (a)  Set I and (b) Set II of the strong-coupling corrections.
The regions of the phase diagram are separated into (A): $v$-vortex is unstable and $d$-vortex is stable, 
(B): $v$-vortex is metastable and $d$-vortex is stable, (C): $v$-vortex is stable and $d$-vortex is metastable. 
The (red) thin-dashed curve represents the phase boundary of the bulk A-phase and bulk planar-phase, which 
is obtained by comparing the free energies Eqs.~\eqref{coreAbulkenergy} 
and \eqref{coreplbulkenergy}. 
}
\label{dvphasedia}
\end{figure}

Let us see the details of the magnetic field dependence of the $v$- and the $d$-vortices in high pressure region.
The free energy of the $v$-vortex increases with $H$, 
while that of the $d$-vortex decreases, as shown in Fig.~\ref{vorenergymag}(d). 
This inverts the energetic stability of the two vortex states at a critical magnetic field. 
Figure \ref{dvphasedia} shows the phase diagram of the stability of the $v$- and the $d$-vortices. 
In addition to the above critical magnetic field, there is another one which gives metastability of the $v$-vortex. 
These critical magnetic fields monotonically increase with the pressure. 

The $p$-dependence of the critical magnetic field can be qualitatively understood by comparing the 
free energy of the bulk A phase and that of the planar phase. The vortex core is filled with these phases. 
For the A phase in the core of the $v$-vortex, described by $\text{Re}(\tilde{A}_{zx}) = \text{Im}(\tilde{A}_{zy}) \equiv \tilde{\Delta}_\text{A1}$, 
the bulk amplitude and the minimized energy are given by
\begin{align}
\tilde{\Delta}_\text{A1}^2 &= \frac{\beta' (1-\tilde{H}^2)}{4 \beta_{245}}, \nonumber \\\
\tilde{f}_\text{A}^H &= - \frac{\beta' (1-\tilde{H})^2}{6 \beta_{245}}.   \label{coreAbulkenergy}
\end{align}
Here, we ignore the minor $\beta$-phase components for simplicity. 
For the planar phase in the $d$-vortex, we take the components $\text{Re}(\tilde{A}_{zx}) \equiv \tilde{\Delta}_\text{p1}$ and 
$\text{Re}(\tilde{A}_{xz}) \equiv -\tilde{\Delta}_\text{p2}$, and the bulk amplitude and the minimized energy are 
\begin{align}
\tilde{\Delta}_\text{p1}^2 &= \frac{\beta'  [\beta_{345} - \tilde{H}^2 (\beta_{12} + \beta_{345})]}{2 \beta_{345} (2 \beta_{12} + \beta_{345})}, \nonumber \\
\tilde{\Delta}_\text{p2}^2 &= \frac{\beta'  (\beta_{12} \tilde{H}^2 + \beta_{345})}{2 \beta_{345} (2 \beta_{12} + \beta_{345})} , \nonumber \\
\tilde{f}_\text{p}^H &= - \frac{\beta'  [\beta_{345} (2- 2 \tilde{H}^2 + \tilde{H}^4) + \beta_{12} \tilde{H}^4]}{6 \beta_{345} (2 \beta_{12} + \beta_{345})} . \label{coreplbulkenergy}
\end{align}
For a given pressure, $\tilde{f}_\text{p}^H$ becomes lower than $\tilde{f}_\text{A}^H$ at a certain magnetic field, 
which is shown by a thin dashed curve in the phase diagram of Fig.~\ref{dvphasedia}. The dashed curve has a similar pressure 
dependence as the critical magnetic fields described above. Thus, we expect that 
the occupation of the planar-phase component is energetically favorable than the A-phase component 
in the vortex core in the presence of a longitudinal magnetic field. 

\subsection{Transverse magnetic field}\label{Hxresults}
We turn to the analysis on the vortices in the transverse magnetic field. Let us consider $\bm{H} = H \bm{e}_x$. 
Then, the magnetic free energy written as $\tilde{f}_\text{M} = \tilde{H}^2 \sum |\tilde{A}_{xi}|^2$ leads to suppression 
of the components $|\tilde{A}_{xi}|$. Here, we should take the boundary condition 
\begin{align}
R(\hat{\bm{x}},\phi) \tilde{A}^\text{(B)} e^{i \theta} = \left( 
\begin{array}{ccc}
\tilde{\Delta}_{\parallel} & 0 & 0  \\ 
0& \tilde{\Delta}_{\perp} \cos \phi & - \tilde{\Delta}_{\perp} \sin \phi \\
0& \tilde{\Delta}_{\perp} \sin \phi & \tilde{\Delta}_{\perp} \cos \phi \\
\end{array}
\right) e^{i \theta}
\label{boundhxcond}
\end{align}
at the boundary $r=R$, where $\tilde{A}^\text{(B)} = \text{diag}(\tilde{\Delta}_{\parallel},\tilde{\Delta}_{\perp},\tilde{\Delta}_{\perp})$ 
with Eq.~\eqref{magampliB}. 
Then, it is not trivial how the rotation angle $\phi$ at the boundary affects the vortex structures, 
because the components $\tilde{A}_{yz}$ and $\tilde{A}_{zy}$ appearing in the vortex cores are influenced by the choice of the angle $\phi$. 
More precisely, when $\phi \neq 0$, not only the components $\tilde{A}_{yy}$ and $\tilde{A}_{zz}$ but also $\tilde{A}_{yz}$ and $\tilde{A}_{zy}$ 
should have zeros around the origin for the nonzero winding number. 
We confirm that the character of all vortices are independent of $\phi$ as shown later; the internal structure of the vortex core
adjusts to the given boundary condition without changing the free energy.

\begin{figure}[ht]
\centering
\includegraphics[width=1.0\linewidth]{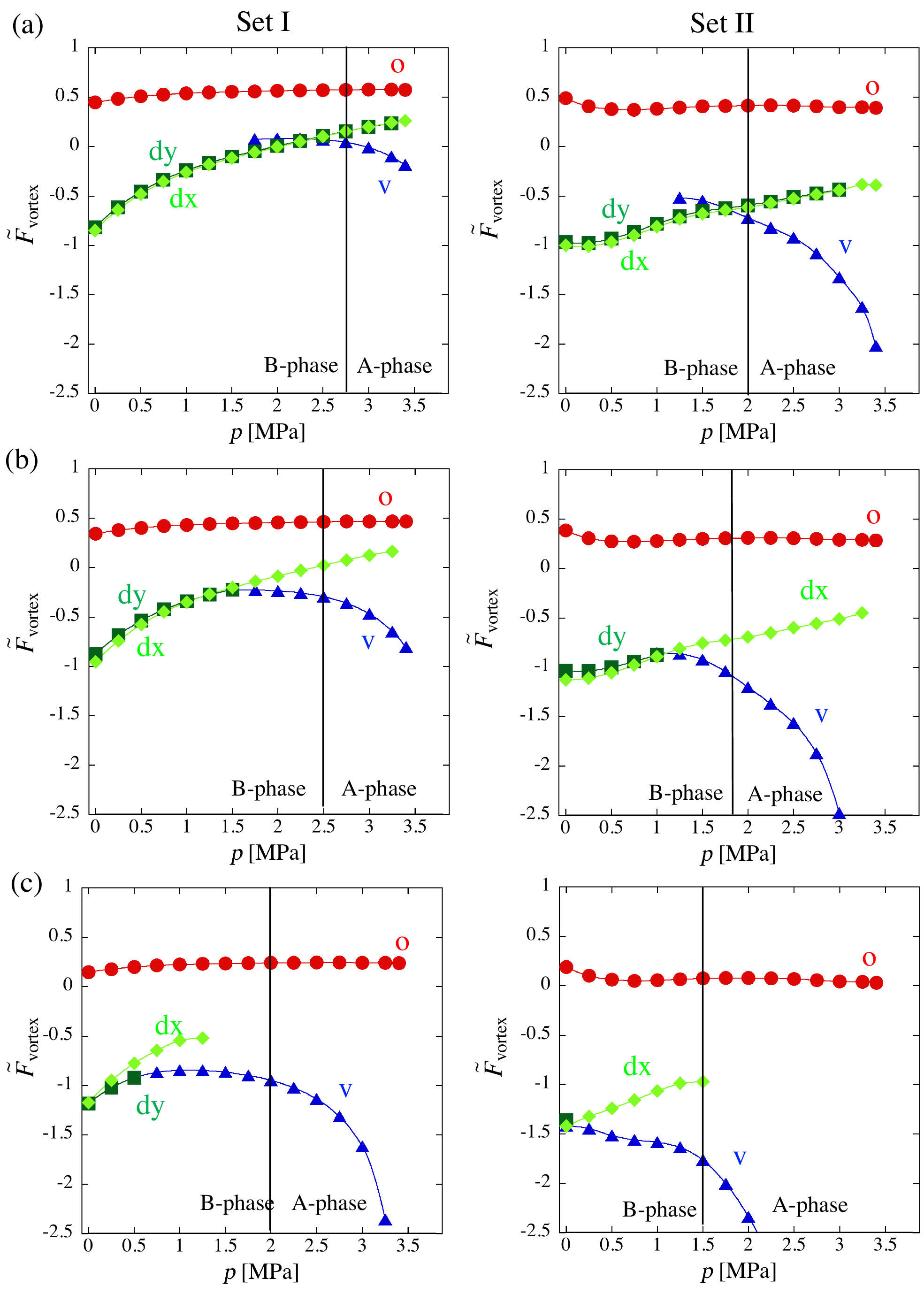} 
\caption{(Color online) Free energies of the vortex states in the presence of a transverse magnetic field as a function 
of the pressure $p$ for Set I (left panels) and Set II (right panels). The magnitude 
of the magnetic field is (a) $\tilde{H} = 0.1$, (b) $\tilde{H} = 0.2$, (c) $\tilde{H} = 0.3$. 
The energies of the $o$-, $v$-, $d_y$- and $d_x$- 
vortices are shown by curves with red circles, blue triangles, green squares, and yellow green diamonds, respectively. The bulk A-B phase 
boundary is taken from Fig.~\ref{phasediabulk}.}
\label{vorenergymaghx}
\end{figure}
\begin{figure}[ht]
\centering
\includegraphics[width=1.0\linewidth]{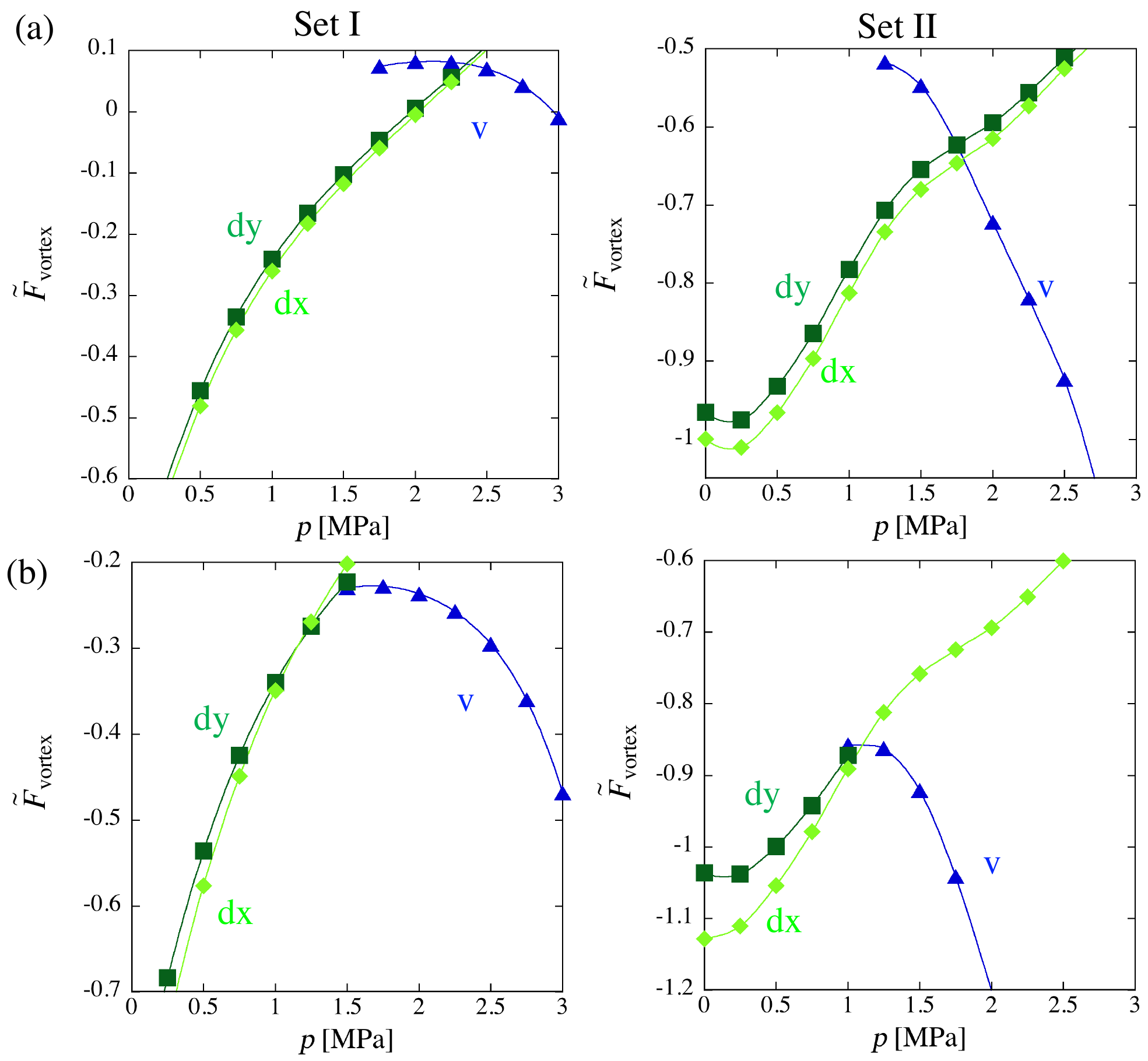} 
\caption{(Color online) Free energies of the $v$-, $d_y$-, and $d_x$-vortex states in the presence of a transverse magnetic field as a function 
of the pressure $p$ for Set I (left panels) and Set II (right panels) for (a) $\tilde{H} = 0.1$ and (b) $\tilde{H} = 0.2$. 
The figures are the enlarged view of Fig.~\ref{vorenergymaghx} (a) and (b) in the vicinity of the transition pressure.}
\label{vorenergymaghxext}
\end{figure}
Here, we analyze the vortex structure by employing the boundary condition Eq.~\eqref{boundhxcond} with $\phi = 0$.
There are two main effects of $\bm{H} = H \bm{e}_x$ on the core structures. One is the stabilization of the $v$-vortex. 
The other is to lock the orientation of the double cores of the $d$-vortex along the $x$- or $y$-direction; we refer to the former and the 
latter as $d_x$-vortex and $d_y$-vortex, respectively. Figure~\ref{vorenergymaghx} shows the vortex energy $\tilde{F}_\text{vortex}$ as a function of 
the pressure, while Fig.~\ref{vorenergymaghxext} depicts a magnified view of Fig.~\ref{vorenergymaghx} in the vicinity of the transition point 
by taking only the energies of the $v$-, $d_x$-, and $d_y$-vortices. 
For $\tilde{H} = 0.1$, the qualitative stability properties of the vortex states are not changed from those for $\tilde{H} = 0$. 
However, the energy of the $d_x$-vortex is slightly lower than that of the $d_y$-vortex. 
With increasing the magnetic field to $\tilde{H} = 0.2$, the stability region of the $v$-vortex extends to the lower pressure region. 
Also, the $d_y$-vortex becomes unstable in the high pressure region, decaying into the symmetric $v$-vortex. 
Above $\tilde{H} = 0.2$, the $d_y$-vortex continuously transforms to the $v$-vortex as $p$ increases. 
No hysteresis is observed in this transition. 
The $d_x$-vortex is stable in the low-pressure region and survives as a metastable state for higher pressures. 
With increasing to $\tilde{H} = 0.3$, although the $d_x$-vortex exists as a metastable state in the low pressure region, the $v$-vortex becomes the most 
stable state. For higher pressures (above 3.25 MPa for Set I and 2.25 MPa for Set II), the vortex state is unstable to become the bulk A-phase without vorticity. 
Similarly to the results in Sec.~\ref{Hzresults}, the $o$-vortex is always metastable.

\begin{figure}[ht]
\centering
\includegraphics[width=1.0\linewidth]{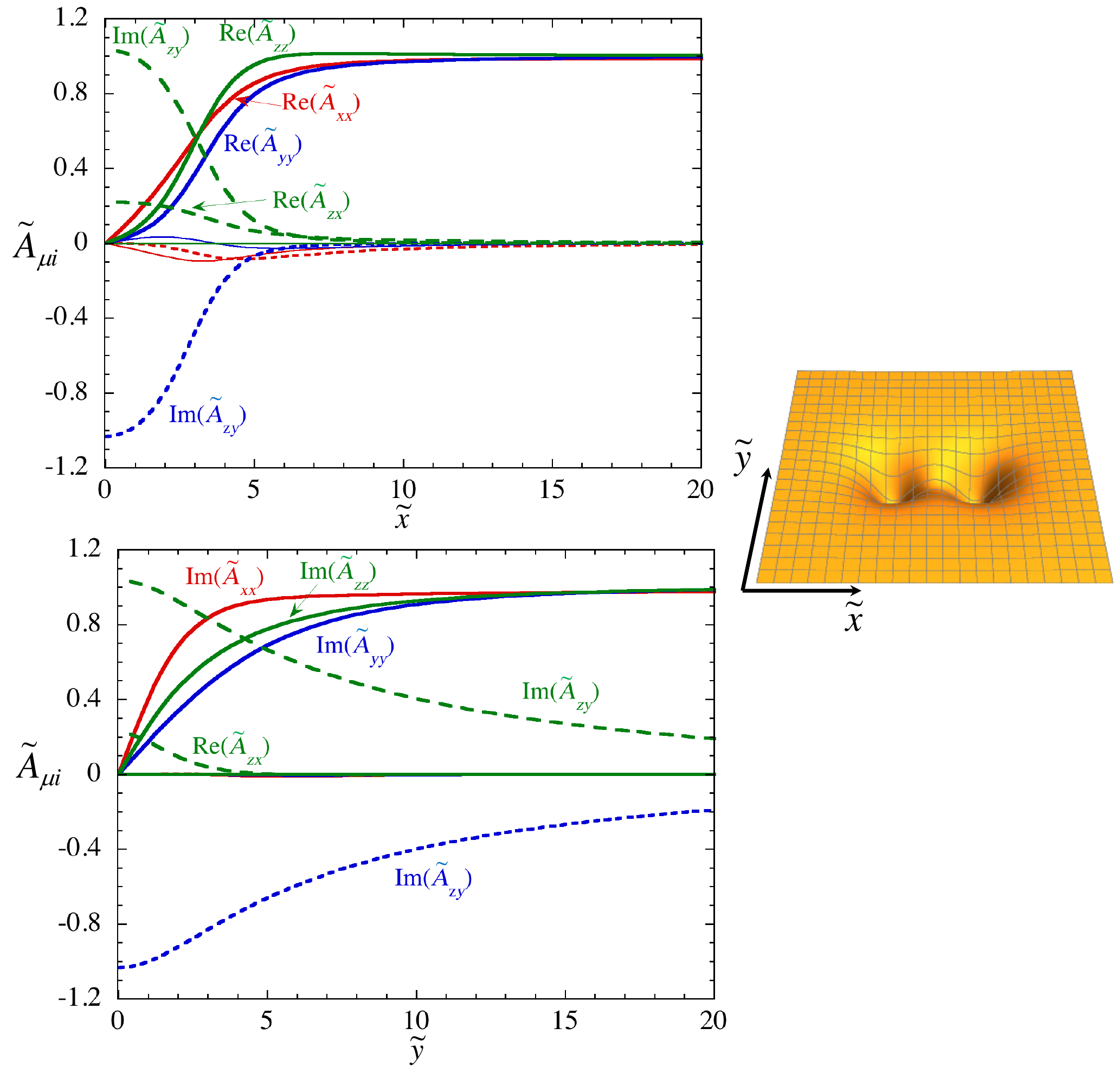} 
\caption{(Color online) Typical radial profiles of the order parameter (left) and the 2D profile of the pair density $\sum_{\mu,i} |\tilde{A}_{\mu i}|^2$ (right) 
of the $d_x$-vortex for $p=$ 3.0 MPa and $\tilde{H} = 0.2$.}
\label{dxvorprof}
\end{figure}
In order to understand the above properties, it is useful to consider the bulk free energy of the core components. The order parameter of the planar phase in the $d$-vortex 
and the rotation angle $\varphi$ of the orientation of the splitting double-core are related with
\begin{align}
\tilde{A}^\text{pl} = 
 \left( 
\begin{array}{ccc}
0 & 0 & - \tilde{\Delta}_\text{p2} a(\varphi) \\
0 & 0 &- \tilde{\Delta}_\text{p2} b (\varphi) \\ 
\tilde{\Delta}_\text{p1} a(\varphi) & \tilde{\Delta}_\text{p1} b (\varphi) & 0
\end{array}\right) 
\end{align}
with $a(\varphi) = \cos^2 \varphi - i \sin \varphi \cos \varphi$ and $b(\varphi) = -\sin \varphi \cos \varphi + i \sin^2 \varphi$. 
Here, $\varphi = 0$ and $\pi/2$ correspond to the $d_y$- and $d_x$-vortices, respectively, where the profiles of the order parameters are 
shown in Fig.~\ref{vorprofilezero}(c) for the $d_y$-vortex and Fig.~\ref{dxvorprof} for the $d_x$-vortex. 
Then, the bulk free energy $f_\text{B}$ is independent of $\varphi$ but the quadratic magnetic free energy for $\bm{H} = H \bm{e}_x$ is written as 
$\tilde{f}_\text{M} = \alpha \tilde{\Delta}_\text{p2}^2 \tilde{H}^2 \cos^2 \varphi$. 
In the bulk, the magnetic free energy is minimized for $\varphi = \pi/2$. 
This feature is consistent with our observation in Fig.~\ref{vorenergymaghxext}. 
We confirm that, through the imaginary time evolution 
from the initial states with $0 < \varphi < \pi/2$, the solutions always converge to those with $\varphi=0$ or $\varphi=\pi/2$. 
This implies that there is an energy barrier between $d_x$- and $d_y$-vortices and the barrier originates from $\tilde{f}_\text{G}$. 

\begin{figure}[ht]
\centering
\includegraphics[width=1.0\linewidth]{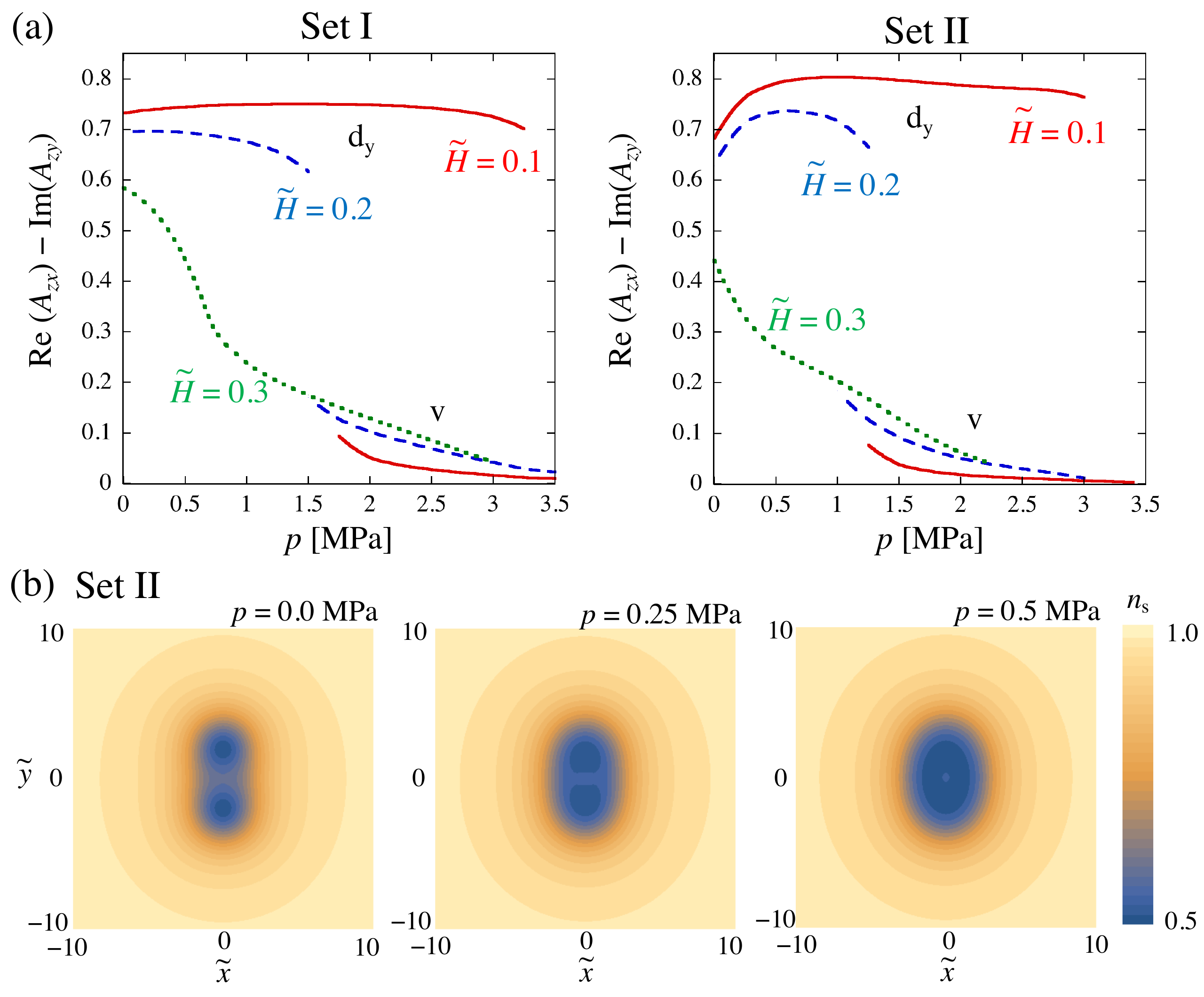} 
\caption{(Color online) Transition from the $d_y$-vortex to the $v$-vortex as $p$ is varied. In (a), the difference Re($\tilde{A}_{zx}$)$-$Im($\tilde{A}_{zy}$) at the origin $x=y=0$ 
is shown as a function of $p$ for $\tilde{H} =$0.1, 0.2, 0.3.  
The left and the right panels correspond to Set I and Set II, respectively. 
The panels in (b) show the contour plots of the pair density $\sum_{\mu,i} |\tilde{A}_{\mu,i}|^2$ for $\tilde{H} = 0.3$ 
and $p=$0, 0.25 and 0.5 MPa from left to right with Set II. }
\label{vorecoraphdif}
\end{figure}
The continuous structural change between the $d_y$-vortex and the $v$-vortex can be analyzed by 
carefully examining the core components. 
For the boundary condition Eq.~\eqref{boundhxcond} with $\phi=0$, the distribution of Re($\tilde{A}_{xz}$) and Re($\tilde{A}_{zx}$) 
in the $v$- and $d_y$-vortices are similar to each other, which allows the continuous transformation between the two structures. 
The angle $\varphi$ of the double-core has to be rotated 
by $\pi/2$ for the $d_x$-vortex to have similar distributions of Re($\tilde{A}_{xz}$) and Re($\tilde{A}_{zx}$) with those in the $v$-vortex. 
This is prohibited by the energy barrier stated above and ensures the presence of the metastable $d_x$-vortex in the high pressure region 
as seen in Figs.~\ref{vorenergymaghx} and \ref{vorenergymaghxext}. The change from the $d_y$-vortex 
to the $v$-vortex can be understood by examining the core components at the origin. 
Figure~\ref{vorecoraphdif}(a) shows the difference Re($\tilde{A}_{zx}$)$-$Im($\tilde{A}_{zy}$) at the origin as a function of the pressure; 
the finite difference implies the $d_y$-vortex with the planar phase core, while the zero difference signifies the $v$-vortex 
with the A-phase core. 
For $\tilde{H} = 0.1$, there is a hysteresis of the transition between the two vortex states. 
As $\tilde{H}$ is increased to 0.2, there appears a jump of the difference without hysteresis. 
For $\tilde{H} = 0.3$, the difference monotonically decreases as $p$ increases, where the continuous change from 
the $d_y$-vortex to the $v$-vortex takes place as $p$ increases, as shown in Fig.~\ref{vorecoraphdif}(b). 
We should note that for $H \neq 0$ the $v$-vortex has the finite difference Re($\tilde{A}_{zx}$)$-$Im($\tilde{A}_{zy}$) in the core components, 
which should be referred to as the \textit{non-axisymmetric $v$-vortex} 

\begin{figure*}[ht]
\centering
\includegraphics[width=1.0\linewidth]{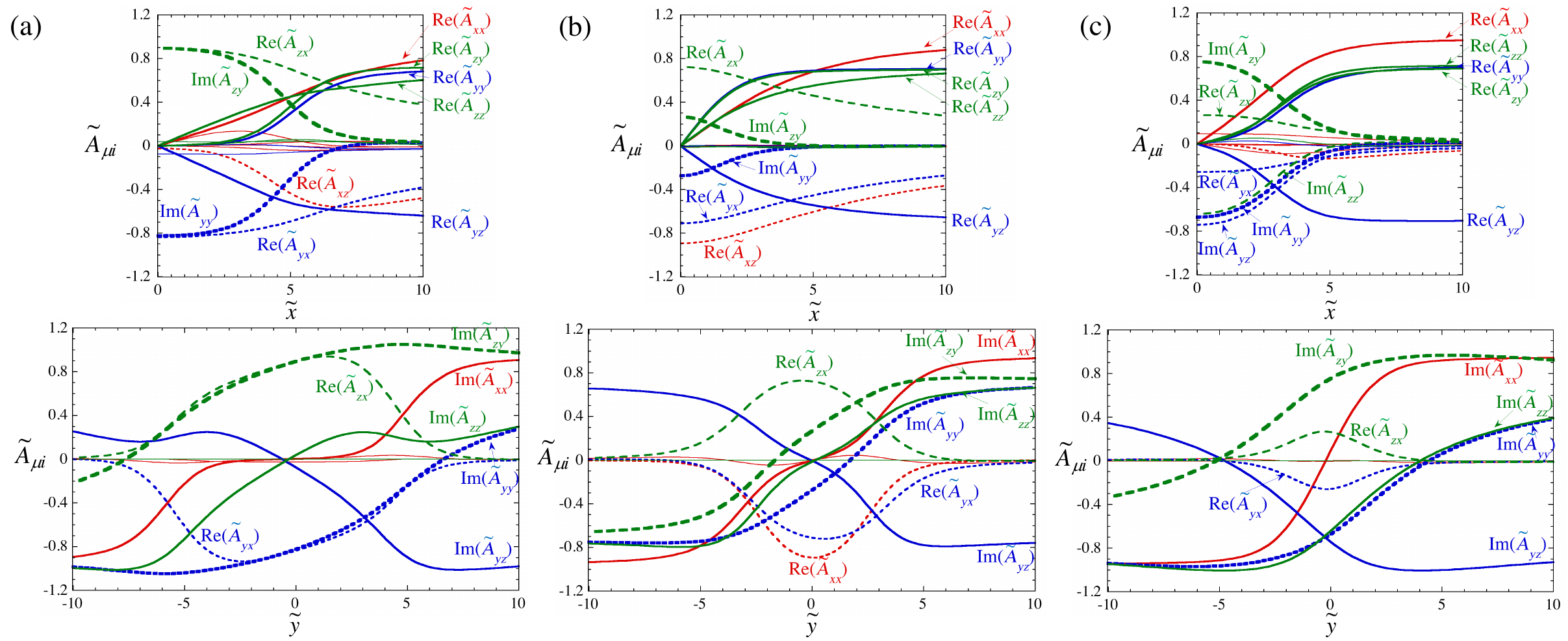} 
\caption{(Color online) The cross section of the order parameter $A$ along the 
$x$- (top) and the $y$- (bottom) axes for (a) the $v$-vortex ($p=3.0$ MPa), (b) the $d_y$-vortex ($p=0.0$ MPa), and (c) the $d_x$-vortex ($p=0.0$ MPa) for $\tilde{H} = 0.2$, 
obtained with the boundary condition Eq.~\eqref{boundhxcond} with $\phi=\pi/4$. 
In all cases, the distributions of the core components along the $x$-axis have inversion symmetry at the origin, 
but those along the $y$-direction do not. }
\label{vorprofileboundary}
\end{figure*}
We checked that the vortex energies are independent of the change of the boundary 
condition associated with the angle $\phi$ of Eq.~\eqref{boundhxcond}. In Fig.~\ref{vorprofileboundary}, we show the profile of the order 
parameters of $v$-, $d_y$-, and $d_x$-vortices for $\phi=\pi/4$. Then, the components $\tilde{A}_{yy}$, $\tilde{A}_{yz}$, $\tilde{A}_{zy}$, $\tilde{A}_{zz}$ 
should have zeros around the origin because of the boundary condition with a phase winding, despite the fact that $\tilde{A}_{yz}$ and $\tilde{A}_{zy}$ should occupy 
the vortex core in each vortex state. Figure \ref{vorprofileboundary} shows that zeros of the order parameter 
appear at the positions displaced from the origin along the $y$-direction. We find numerically that the total energy as well as 
the total pair density $\sum_{\mu,i} |\tilde{A}_{\mu,i}|^2$ are invariant with respect to $\phi$. 

\begin{figure}[ht]
\centering
\includegraphics[width=1.0\linewidth]{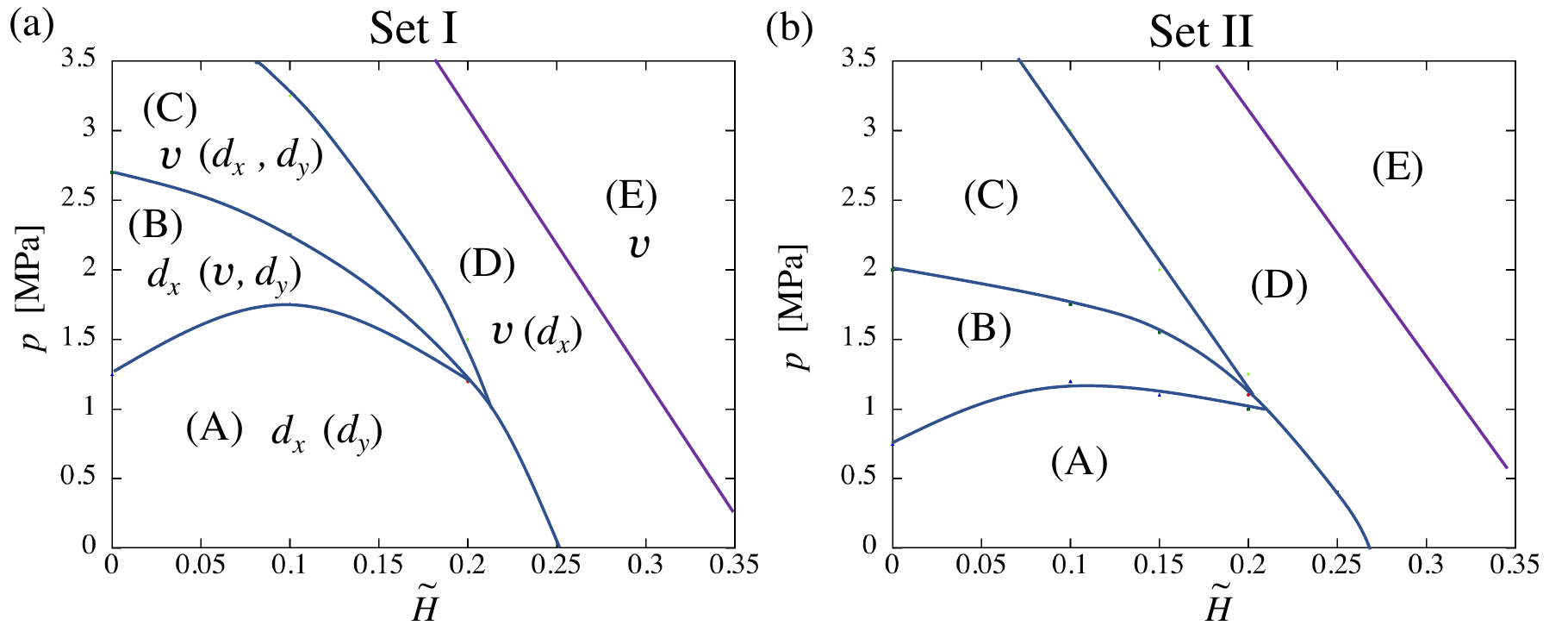} 
\caption{(Color online) $p$-$H$ phase diagram of the stable $v$- and $d$-vortex states 
for (a)  Set I and (b) Set II of the strong-coupling corrections.
The $p$-$H$ plane is divided into 
(A): the $d_x$-vortex is stable and the $d_y$-vortex is metastable, 
(B): the $v$-vortex is metastable, the $d_x$-vortex is stable, and the $d_y$-vortex is metastable, 
(C): the $v$-vortex is stable and the $d_x$- and $d_y$-vortices are metastable, 
(D): the $v$-vortex is stable and the $d_x$-vortex is metastable, and 
(E): the $v$-vortex is stable. }
\label{dvphasehx}
\end{figure}

Finally, our results are summarized in the $p$-$H$ phase diagram of Fig.~\ref{dvphasehx}. 
As an overall feature, with increasing the transverse magnetic field $\tilde{H}$, the stable region of the $d$-vortex 
gradually shrinks, while that of the $v$-vortex gradually expands. 

\section{conclusion}\label{concle}
In this paper, we study the vortex states in superfluid $^3$He-B phase under a longitudinal and a transverse magnetic fields within the 
GL theory. We calculate the free energy of the $o$-, the $v$- and the $d$-vortices with two different 
sets of the strong-coupling corrections, Set I and Set II, finding that the results for Set II reproduce quantitatively the 
experimental data at zero magnetic fields. Under a magnetic field along the $z$-axis (the axis of the vortex), 
the $v$-vortex, which is stable at high pressure and zero magnetic field, becomes unstable and decays 
to the $d$-vortex. 
On the other hand, if the magnetic field is applied along the $x$-axis, perpendicular to the axis of the vortex, 
the $v$-vortex is the most stable state at high magnetic field. 
In the latter case, anisotropic feature of the vortex structure is more remarkable, where the $d_x$- and the $d_y$-vortices 
exist as distinct energy minima. We estimate that these transitions take place at $H \sim 0.2 H_{0} \sim 50$ mT, 
when $T = 0.9 T_c$, which is observable under suitable experimental setup.\cite{Krusiusrev,Hakonen,Kondo} 

The dependence of the vortex structure on the magnetic field can be understood qualitatively by the 
energetic argument of the core-occupying components of the order parameter. 
A magnetic field applied to a bulk superfluid $^3$He stabilizes the A-phase more than the B-phase in general.
The $d$-vortex tends to be stabilized in the presence of a magnetic field along the $z$-axis. 
Recall that the core of the $v$-vortex is filled with the A-phase with non-vanishing order parameter components 
$A_{zi} (i=x,y)$, which are suppressed by the magnetic field along the $z$-axis. 
This makes the $v$-vortex energetically unfavorable compared to the $d$-vortex. 
This is rather counterintuitive since a magnetic field stabilizes the bulk A-phase. 
A magnetic field along the $x$-axis suppresses the order parameter components $A_{xi} (i=x,y,z)$. 
The core of the $v$-vortex is filled with the components $A_{zx}$ and $A_{zy}$ while that of the $d_x$-vortex, 
the stable $d$-vortex in this case, is filled with the planar phase with dominant components Im$A_{yz}$ and Im$A_{zy}$. 
These core components are not suppressed by the transverse magnetic field. However, the $v$-vortex is more stable 
than the $d_x$-vortex when the transverse magnetic field is strong enough since a magnetic field energetically 
favors the A-phase compared to the planar phase for a bulk superfluid $^3$He. 
As a further study, it is interesting to see the effect of a magnetic field applied obliquely to the vortex axis.

\acknowledgements
We would like thank Erkki Thuneberg and Matti Krusius for their interest in this work.
The work of K.K. is partly supported by KAKENHI from the Japan Society for the Promotion of Science (JSPS) Grant-in-Aid for Scientific Research (KAKENHI Grant No. 18K03472). 
The work of M.N. is partly supported by KAKENHI from the JSPS Grant-in-Aid for Scientific Research (KAKENHI Grant No. 17K05554).

\end{document}